# Level correlations driven by weak localization in 2-d systems

Vladimir E. Kravtsov[1] and Igor V. Lerner[2]

[1] *International Centre for Theoretical Physics, 34100 Trieste, Italy*
*and Institute of Spectroscopy, Russian Academy of Sciences, 142092 Troitsk, Moscow r-n, Russia*
[2] *School of Physics and Space Research, University of Birmingham, Edgbaston, Birmingham B15 2TT, United Kingdom*

We consider the two-level correlation function in two-dimensional disordered systems. In the non-ergodic diffusive regime, at energy $\varepsilon > E_c$ ($E_c$ is the Thouless energy), it is shown to be completely determined by the weak localization effects, thus being extremely sensitive to time-reversal and spin symmetry breaking: it decreases drastically in the presence of magnetic field or magnetic impurities and changes its sign in the presence of a spin-orbit interaction. In contrast to this, the variance of the levels number fluctuations is shown to be almost unaffected by the weak localization effects.
(Submitted to Phys.Rev.Lett. 19 Septmeber 1994)

A remarkable observation of Wigner and Dyson [1] was the existence of universal statistics that govern energy level spectra in a wide variety of quantum systems [2]. The Wigner–Dyson statistics of eigenvalues in the invariant ensembles of random matrices are applicable also to disordered electron systems in a metallic phase [3–6]. The *level repulsion*, characteristic of these statistics, results from the overlaping of one-electron states corresponding to different energies. With the disorder increasing such systems undergo the Anderson metal-insulator transition [7]. On an insulator side of the transition, uncorrelated energy levels are described by the Poisson statistics.

Recently, the existence of the third universal spectral statistics, applicable to the disordered systems in a critical regime near the Anderson transition point, has been predicted [8–11]. These statistics emerge [10] in the crossover region near the transition, where electron diffusion is anomalous, and turn out to be completely new [9] and different from the Poisson and Wigner-Dyson ones, rather than a universal hybrid of both, as has been earlier conjectured [8]. They are still characterized by the level repulsion [11], albeit it is weaker than in the metallic region. It is essential that nontrivial level correlations [9,10] turn out to be driven by small scale-dependent *corrections* to the conductance which by itself is almost scale-independent near the Anderson transition.

The conductance has only a weak scale-dependence also in the case of a two-dimensional disordered system in a weak-localization (WL) regime. There is no metal-insulator transition at $d=2$, and, with the size increasing, such a system crosses over to a strong-localization regime [12]. So one can expect some analogies between the level correlations in this case and those in the critical regime in the higher dimensionality. However, while the one-parameter scaling hypothesis [12] was taken for granted to make considerations in the critical regime possible [9,10], the WL regime at $d=2$ may be treated within a rigorous and conjecture-free perturbative approach.

In this Letter we consider level correlations in a $2d$ disordered electron system and show that they are totally governed by the WL corrections to the conductance, when the distance between the levels, $\varepsilon$, is in the region

$$E_c \lesssim \varepsilon \lesssim \hbar/g\tau \qquad (1)$$

This is another of a few known examples (as an anomalous magnetoresistance or the Aharonov-Bohm effect [7]) where the weak-localization *corrections* determine the *main effect*. In Eq. (1), $E_c = \hbar D_0/L_0^2$ is the Thouless energy, $L_0$ is the sample size, $D_0 = \ell v_F/2$ is the diffusion coefficient, $\ell = v_F \tau$ is the elastic scattering length, and $g$ is the conductance in units of $e^2/\pi\hbar$.

We consider the two-level correlation function:

$$R_E(\varepsilon) = \frac{\langle \rho(E)\rho(E+\varepsilon)\rangle}{\langle \rho(E)\rangle\langle \rho(E+\varepsilon)\rangle} - 1. \qquad (2)$$

Here $\langle \ldots \rangle$ denote averaging over all the realizations, and

$$\rho(E) = pL^{-d}\sum_n \delta(E - \varepsilon_n) \qquad (3)$$

is an exact density of states for a particular realization of disorder, where $\varepsilon_n$ are exact energy levels, and $p$ is a spin degeneracy. For $E$, $E+\varepsilon$ in the bulk of conduction band, $\langle \rho(E)\rangle \equiv \rho$ is energy-independent, and $R_E(\varepsilon) \equiv R(\varepsilon)$ is a function of $(\varepsilon)$ only.

The averaging in Eq.(2) is performed within the impurity diagrammatic technique [13] with $\rho(E)$ represented by $(ip/\pi L^d)\int \mathrm{Im}\,G^+(\mathbf{r},\mathbf{r};E)d\mathbf{r}$ (here $G^\pm$ are exact retarded or advanced one-particle Green's functions). It is convenient to use a representation in which slow diffusion modes are explicitly separated from fast "ballistic" ones [14,15]. In Fig. 1, there are shown the lowest order (one-loop) diagrams containing one or two diffusion or cooperon propagators (shown by wavy lines which correspond to the ladder series in the conventional technique [5]). Triangular and square "Hikami boxes" represent the motion at the ballistic scale as all the averaged Green's functions $\overline{G}^\pm(\mathbf{r}-\mathbf{r}',\varepsilon)$ (represented by edges of the boxes) decay as $e^{-|\mathbf{r}-\mathbf{r}'|/2\ell}$. At the diffusion scale, these boxes are reduced to certain constants (see Ref. [16] for review). Taking into account the total number of cooperon and diffuson contributions that depends on the universality class, one obtains for the one- and two-diffuson diagrams:

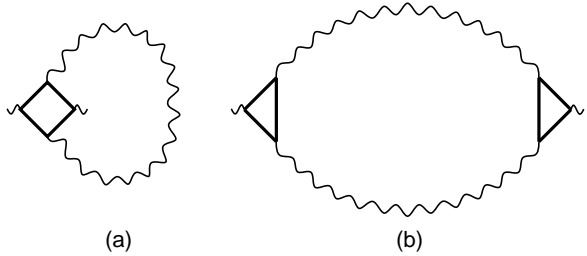

FIG. 1. One-loop diagrams contributed to $R(\varepsilon)$.

$$R_{(1)}(\varepsilon) = \frac{2\Delta^2 \tau}{\pi^2 \beta \hbar} \mathrm{Re} \sum_q \frac{1}{(\hbar D_0 q^2 - i\varepsilon)}, \quad (4a)$$

$$R_{(2)}(\varepsilon) = \frac{\Delta^2}{\pi^2 \beta} \mathrm{Re} \sum_q \frac{1}{(\hbar D_0 q^2 - i\varepsilon)^2}. \quad (4b)$$

where $q = (2\pi/L_0)(n_x, n_y)$, and $n_{x,y}$ run over all the integers, $\beta = 1$, 2, or 4 is for the Dyson orthogonal, unitary, and symplectic ensembles, respectively, $\Delta = p/\rho L^d$ is the mean level spacing.

The two-diffuson diagram of Fig. 1b is well-known to dominate at $d > 2$ in the interval $\Delta \lesssim \varepsilon \lesssim \hbar/\tau$. Indeed, in the ergodic region, $\varepsilon \ll E_c$, where all the terms with $q \neq 0$ may be neglected in Eq. (4b), it gives

$$R(\varepsilon) \simeq -\frac{1}{\pi^2 \beta s^2}, \quad s \equiv \frac{\varepsilon}{\Delta}, \quad (5)$$

which is the envelope of the Wigner–Dyson correlation function. The contribution of the one-diffuson diagram, Eq. (4a), is of order $(l/L_0)^d (lk_F)^{-2(d-1)}$ in this region, i.e. totally negligible. In the diffusive region, $E_c \ll \varepsilon \ll \hbar/\tau$, the summation in Eqs. (4) may be changed by integration, and the two-diffuson diagram, Eq. (4b), gives the well-known result [5]:

$$R(\varepsilon) = \frac{\Delta^2 L^d}{\pi^2 \beta (2\pi)^d} \mathrm{Re} \int \frac{d^d q}{(\hbar D_0 q^2 - i\varepsilon)^2} = \frac{C_d}{\beta g^{d/2} |s|^{2-\frac{d}{2}}}. \quad (6)$$

Here the numerical coefficient $C_d$ depends only on $d$.

For $d > 2$, Eq. (6) makes the main contribution into $R(\varepsilon)$, as that of the diagram of Fig. 1a turns out to be smaller by $(l/L_\varepsilon)^{4-d}$, where $L_\varepsilon^2 = \hbar D_0/\varepsilon$. It becomes relevant for the correlation function (2) only at the ballistic energy scale $\varepsilon \sim \hbar/\tau$, where the appropriate diffusion length $L_\varepsilon$ is of order $\ell$ and the diffusion approximation is no longer valid.

For $d = 2$ such an estimation fails as $C_2 = 0$ in Eq.(6). A similar cancellation (of a contrubution that would be main if not a vanishing coefficient) happens at the mobility edge for $d > 2$, although much more subtle considerations were required [9]. On the face of it, what is left at $d = 2$ is the contribution of the diagram of Fig. 1a and the next order correction for the diagram of Fig. 1b (arisen when the corrections to the boxes $\propto (q\ell)^2$ are taken into account). It gives:

$$R_{\mathrm{bal}}(\varepsilon) = \frac{\tau \Delta}{2\pi \beta g \hbar} \ln\left[\frac{(\varepsilon \tau/\hbar)^2}{1 + (\varepsilon \tau/\hbar)^2}\right]. \quad (7)$$

This result corresponds to that obtained by Altland and Gefen [17] who have thoroughly considered the level correlations in the ballistic regime (in their technique both the diagrams of Fig. 1 were described by a single expression so that no explicit cancellation has been seen). In the diffusive regime, it gives only a logarithmic dependence on $\varepsilon$ with a small amplitude, $\tau \Delta/g \hbar \sim (1/k_F L_0)^2$.

A surprising situation at $d = 2$ is that in a wide region, Eq. (1), the contribution of the *one-loop diagrams*, Eq. (7), is small compared to that of the *two-loop diagram* containing four propagators (Fig. 2a). The shaded Hikami box is $\propto q^2$ [15], so that this diagram describes the WL correction to the diffusion coefficient $D_0$ in Eqs. (4). It contains a cooperon, thus being absent in the $\beta = 2$ case where the next-order (three-loop) diagrams should be considered. Therefore, to find the WL contribution to the correlation function (2), we calculate the correction to $D_0$, given for $\beta = 1, 4$ by

$$\delta D(\varepsilon) = \frac{\Delta}{\pi \hbar (\beta - 2)} \sum_q \frac{1}{q^2 - iL_\varepsilon^{-2}}, \quad (8)$$

and substitute $D = D_0 + \delta D$ into the integral in Eq. (6). Naturally, its real part would not change the zero result at $d = 2$. So it is the *imaginary part* of the WL correction that governs the level correlations in the region (1). With the same accuracy as in Eq. (6),

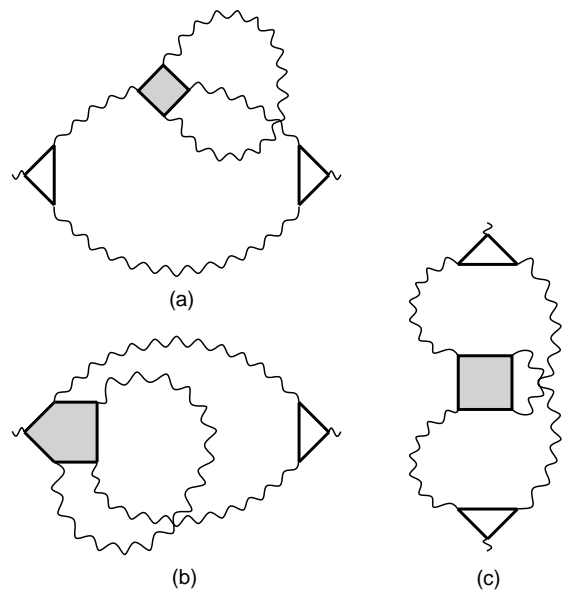

FIG. 2. Two-loop diagrams contributed to $R(\varepsilon)$. The diagrams (b) and (c), that describe renormalization of the effective vertex, cancel each other. The diagram (a) yields the WL correction to the diffusion coefficient.

$$\operatorname{Im}\delta D(\varepsilon) = \frac{i\eta(g,\beta)\,\operatorname{sign}(\varepsilon)}{4\pi\hbar\rho}. \tag{9}$$

Here $\eta(g,\beta) = (\beta-2)^{-1}$ for $\beta = 1, 4$, and $= -1/2g$ for $\beta = 2$ where the above result is obtained with due regard to the third-loop corrections, and thus contained an additional factor of $1/g$. Calculating the integral (6) with allowance for the correction (9), we obtain

$$R_{\mathrm{WL}}(\varepsilon) = \frac{\eta(g,\beta)}{\beta}\frac{1}{g^2}\frac{1}{|s|}. \tag{10}$$

There exist two more two-loop diagrams, Fig. 2b,c, but they exactly cancel each other. It has been expected, as these diagrams would yield a singular correction to a renormalized triangular vertex of the diagram of Fig. 1b. The vertex has zero scaling dimensionality, and should remain unrenormalized, as its renormalization would violate the particle conservation law. Thus all the vertex corrections must cancel each other, as has been noted in Ref. [9] for the diagrams at the mobility edge. The cancellation of the diagrams in Fig. 2b,c provides a perturbative proof of the absence of the vertex renormalization.

In the region (1) this contribution dominates over that of Eq. (7), provided that $g < \hbar/E_c\tau \sim (L_0/\ell)^2$. (For $\beta=2$, it dominates for $E_c\varepsilon \lesssim \hbar/g^2\tau$, such region existing for $g \lesssim L_0/\ell$). This condition may be violated for a very small or a very pure sample which is almost ballistic. In general, however, it is easily satisfied. The two contributions are matching for $\varepsilon \sim \hbar\eta/g\tau$, and the latter becomes dominating for the higher energies, in the 'quasi-ballistic' region $\hbar\eta/g\tau \lesssim \varepsilon \lesssim \hbar/\tau$. There is, however, a parametric mismatching of the results of Eqs. (5) and (10) at the boundary between the ergodic and diffusive regime, $\varepsilon \sim E_c$, where the ergodic expression, Eq. (5), is of order $-1/g^2$ while diffusive one is of order $-1/g^3$ for $\beta=1$, $-1/g^4$ for $\beta=2$, and $+1/g^3$ for $\beta=4$. The reason is that one can justify neglecting all the $q \neq 0$ terms in the sum (4b) only for $\varepsilon \ll E_c$, and changing summation by integration only for $\varepsilon \gg E_c$. There exists some transient region near $\varepsilon \sim E_c$ where $R(\varepsilon)$ crosses over from the Wigner-Dyson behavior, Eq. (5), to the WL behavior, Eq. (10). In this region one should evaluate accurately the sum in Eq. (4b). Changing there to the dimensionless variables $n$ and $s$, and using the transformation $b^{-2} = \int_0^\infty \alpha e^{-\alpha b}d\alpha$, one performs the summation over $n_x$ and $n_y$ in Eq. (4b) to obtain the result in terms of the elliptic theta-function [18] $\theta_3$:

$$R_{(2)}(s) = \frac{1}{\pi^2\beta g^2}\,\operatorname{Re} f(s), \tag{11a}$$

$$f(s) = \sum_{n_x,n_y}\frac{1}{(n^2-is/g)^2} = \int_0^\infty \alpha e^{i\alpha s/g}\varphi^2(\alpha)d\alpha, \tag{11b}$$

$$\varphi(\alpha) = \sum_{-\infty}^{+\infty} e^{-\alpha n^2} = \theta_3(0, e^{-\alpha}). \tag{11c}$$

The asymptotical values of $\varphi$ are given by

$$\varphi(\alpha) = \begin{cases} \sqrt{\pi/\alpha}\,[1 + 2\exp(-\pi^2/\alpha)], & \alpha \ll 1; \\ 1 + 2e^{-\alpha}, & \alpha \gg 1. \end{cases} \tag{12}$$

For $s \ll g$, Eq. (11a) is thus reduced to the ergodic result (5), while for $s \gg g$, $\operatorname{Re} f(s) \propto \exp[-\pi\sqrt{2s/g}]$. Therefore, for large $s$ using the accurate summation instead of integration gives only the exponentially small correction to the WL result of Eq. (10). (Note that with the accuracy up to the same exponent, $\operatorname{Im} f(s) = \pi/s$, so that using the integration for calculating the WL correction is justified). In the transient region of the width $\varepsilon_\circ = 2E_c\ln^2 g$, the contribution of $\operatorname{Re} f(s)$ is important (and dominant for $|\varepsilon - E_c| \ll \varepsilon_\circ$), and it describes a smooth and *universal* crossover from the Wigner-Dyson to the WL regime in $R(s)$.

Now we can combine the results in the ergodic region, Eq. (5), the WL region, and the transient region described above to represent them as

$$R(s) = \frac{1}{\pi^2\beta g^2}\left[\operatorname{Re} f(s) + \frac{\pi^2\eta}{|s|}\right] \tag{13}$$

Only for $s \gtrsim (\hbar/g\tau\Delta)$, this contribution becomes smaller than the quasi-ballistic one, Eq. (7), found in Ref. [17].

The two-level correlation function $R(s)$ studied above determines the variance $\Sigma_2 = \langle(N-\overline{N})^2\rangle$ of the number of levels $N$ in an energy strip of the given width $E = \overline{N}\Delta$ (here $\langle N \rangle \equiv \overline{N}$). Its derivative is given by

$$\frac{d\Sigma_2}{d\overline{N}} = \int_{-\overline{N}}^{\overline{N}} R(s)ds. \tag{14}$$

In order to calculate the r.h.s of Eq.(14), it is convenient to use the representation of $R(s)$ in the form of a sum, as in Eq. (11b), and integrate first over $s$. Then, neglecting the WL corrections, one obtains for $\overline{N} \gg g$:

$$\frac{d\Sigma_2}{d\overline{N}} = \frac{2\overline{N}}{\pi^2\beta g^2}\sum_n \frac{1}{[(\overline{N}/g)^2 + (n^2)^2]} \simeq \frac{1}{\beta g}, \tag{15}$$

so that $\Sigma_2 \sim \overline{N}/g$ [5]. Allowing for the WL term, one finds with the logarithmic accuracy:

$$\Sigma_2(E) = \frac{\overline{N}}{\beta g}\left[1 + \frac{2\eta(g,\beta)}{g}\ln\left(\frac{\overline{N}}{g}\right)\right]. \tag{16}$$

As in the WL regime $g^{-1}\ln(L_0/l) \ll 1$, the variance $\Sigma_2$ is almost linear in $\overline{N}$ in the whole region $E_c < E = \overline{N}\Delta < \hbar/\tau$. Indeed, the WL logarithmic term in Eq.(16) could be important only for $\overline{N} \gtrsim ge^g$. For $E \lesssim \hbar/\tau$, this gives $e^g \lesssim \hbar/E_c\tau \sim (L_0/l)^2$ while the WL approach is applicable in the opposite limit, $g^{-1}\ln(L_0/l) \lesssim 1$.

Note that the $\overline{N}$-independent result of Eq. (15) is only valid in the diffusive region $\hbar/(\tau\Delta) \gg \overline{N} \gg g$. For $\overline{N} \to \infty$ the integral in the r.h.s. of Eq. (14) *must vanish*. The reason is that there exists a sum rule which follows from the conservation of the total number of states $\mathcal{N}$.

Multiplying Eq. (2) by $\langle \rho(E+\varepsilon) \rangle$ and integrating over all $\varepsilon$ with the identity $L^d \int \rho(E+\varepsilon)\,d\varepsilon = \mathcal{N}$, one obtains:

$$\int R_E(\varepsilon)\langle \rho(E+\varepsilon)\rangle\,d\varepsilon = \int_{-\infty}^{+\infty} R(s)\,ds = 0, \qquad (17)$$

where the second identity in Eq.(17) results from $\langle \rho \rangle$ being a constant in the relevant interval, $\varepsilon \ll \varepsilon_F$.

By the derivation, the sum rule Eq.(17) holds exactly as long as the total number of states $\mathcal{N}$ is finite. However, it may be violated if the thermodynamic (TD) limit, $\mathcal{N} \to \infty$, were taken before the integration in Eq.(17).

To illustrate this point let us consider an ensemble of *diagonal* random $\mathcal{N} \times \mathcal{N}$-matrices. Suppose that matrix elements, $x_n$, are independent random variables each having the same probability distribution: $P(x) = \mathcal{N}^{-1}$ for $|x| < \mathcal{N}/2$ and $P(x) = 0$ otherwise. Using the definitions (2, 3), one immediately finds the correlation function, $R_0(s) = \delta(s) - \mathcal{N}^{-1}\theta(\mathcal{N}/2 - |s|)$, where $\theta(x)$ is a step function. For finite $\mathcal{N}$, $R_0(s)$ obeys the sum rule (17), since a positive contribution of the $\delta$-function is cancelled by a negative one given by the flat small "tail" in $R_0(s)$. The universal Poisson distribution, $R_0(s) = \delta(s)$, arises only in the TD limit and obviously does not obey the sum rule.

The above toy model illustrates what happens in the present problem. The integral in Eq. (14) is really vanishing in the limit $\overline{N} \to \infty$, as required by the sum rule, due to the negative contribution, $(-\beta g)^{-1}$, of the long and small "ballistic" tail in $R(s)$ given by Eq. (7) for $\varepsilon \tau/\hbar \sim 1$. If the TD limit, $L_0 \to \infty$, is taken in $R(s)$ this ballistic tail vanishes, and the sum rule breaks down.

A situation similar to the present $2d$ case takes place at $d > 2$ at the mobility edge where the sum rule for the limiting function $R(s)$ breaks down for the same reason. Thus, in contrast to the statement made in Ref. [9], it could not be applied to calculating the variance. This leads to a linear in $\overline{N}$ term in the variance [19], missed in Ref. [9], which is similar to that in Eq. (16). A difference is that $g \sim 1$ at the mobility edge so that the coefficient of proportionality is just a certain number. The long-range levels correlations lead to the contribution to the variance $\propto \overline{N}^\gamma$ ($\gamma$ is a universal critical exponent), so that $\Sigma_2 = AN + BN^\gamma$ where $A$ and $B$ are contributed by *all* the diagrams and thus cannot be exactly determined.

We conclude that in the diffusive region of Eq. (1) the level correlations, Eq. (10), are totally *governed* by the weak localization effects while the level number variance, Eq. (16), is governed by the *total* conductance with the WL effects resulting only in small corrections.

## ACKNOWLEDGMENTS

We are grateful to A. Aronov, D. Khmelnitskii, G. Montambaux, B. Shapiro, and B. Shklovskii for useful discussions on this and related subjects. V.E.K. is thankful to the University of Birmingham, and I.V.L. is thankful to the INT at the University of Washington, Seattle, for kind hospitality extended to them at different stages of this work. The partial support of the EEC grant CT90 0020 (V.E.K. and I.V.L.) and the NATO grant CRG.921399 (I.V.L.) is gratefully acknowledged.